\documentclass[10pt,conference]{IEEEtran}

\usepackage{amsmath,amssymb}
\usepackage{graphicx,graphics,color,psfrag}
\usepackage{cite,balance}
\usepackage{algorithm}
\usepackage{accents}
\usepackage{bm}
\usepackage{url}
\usepackage{algorithmic}
\usepackage[english]{babel}
\usepackage{multirow}
\usepackage{enumerate}
\usepackage{cases}
\usepackage{stfloats}
\usepackage{dsfont}
\usepackage{color,soul}
\usepackage{amsfonts}
\usepackage{tcolorbox}
\usepackage{amsmath}
\usepackage{float}
\usepackage{mathtools}
\usepackage{verbatim} 
\usepackage{bm,color,soul}
\usepackage{epsfig}
\usepackage{subfigure}
\usepackage{psfrag}

\usepackage{latexsym}

\usepackage{color}
\usepackage{booktabs}

\IEEEoverridecommandlockouts

\usepackage{algorithm}
\usepackage{algorithmic}

\usepackage{cite,graphicx,amsmath,amssymb}
\usepackage{fancyhdr}
\usepackage{hhline}
\usepackage{graphicx,graphics}
\usepackage{array,color}
\usepackage{amsmath}
\usepackage{stfloats}
\usepackage[flushleft]{threeparttable}
\usepackage{makecell} 

\begin{document}

%
\title{Coordinated Transmit Beamforming for Multi-antenna Network Integrated Sensing and Communication}

\author{
\IEEEauthorblockN{Gaoyuan~Cheng and  Jie~Xu%
\thanks{ J. Xu is the corresponding author. }
}

\IEEEauthorblockA{School of Science and Engineering (SSE) and Future Network of Intelligence Institute (FNii), \\ The Chinese University of Hong Kong (Shenzhen),  China}
Email: gaoyuancheng@link.cuhk.edu.cn, xujie@cuhk.edu.cn
}

\maketitle

\begin{abstract}
This paper studies a multi-antenna network integrated sensing and communication (ISAC) system, in which a set of multi-antenna base stations (BSs) employ the coordinated transmit beamforming to serve their respectively associated single-antenna communication users (CUs), and at the same time reuse the reflected information signals to perform joint target detection. In particular, we consider two target detection scenarios depending on the time synchronization among BSs. In Scenario \uppercase\expandafter{\romannumeral1}, these BSs are  synchronized and can exploit the target-reflected signals over both the direct links (from each BS to target to itself) and the cross links (from each BS to target to other BSs) for joint detection. In Scenario \uppercase\expandafter{\romannumeral2}, these BSs are not synchronized and can only utilize target-reflected signals over the direct links for joint detection. For each scenario, we derive the detection probability under a specific false alarm probability at any given target location. Based on the derivation, we optimize the coordinated transmit beamforming at the BSs to maximize the minimum detection probability over a particular target area, while ensuring the minimum signal-to-interference-plus-noise ratio (SINR) constraints at the CUs, subject to the maximum transmit power constraints at the BSs. We use the semi-definite relaxation (SDR) technique to obtain highly-quality solutions to the formulated problems. Numerical results show that for each scenario, the proposed design achieves higher detection probability than the benchmark scheme based on communication design. It is also shown that the time synchronization among BSs is beneficial in enhancing the detection performance as more reflected signal paths are exploited.
\end{abstract}


\section{Introduction}


Integrated sensing and communication (ISAC) has been recognized as an enabling technology towards future sixth-generation (6G) wireless networks to support new applications such as auto-driving, smart city, and industrial automation \cite{liu2022integrated,Liu2022A}. On one hand, ISAC allows to share cellular infrastructures like base stations (BSs) as well as scarce spectrum and power resources for the dual roles of communication and sensing, thus increasing the resource utilization efficiency. On the other hand, ISAC enables the joint sensing and communication optimization within the integrated system, thus helping better manage their co-channel interference and accordingly enhancing the system performance.

Conventionally, mono-static and bi-static ISAC systems have been widely investigated in the literature (see, e.g., \cite{kumari2019adaptive,hua2021optimal} and the references therein), in which one BS acts as an ISAC transceiver or two BSs serve as the ISAC transmitter and the sensing receiver, respectively. However, such ISAC systems normally have limited coverage, and the resulting sensing and communication performances may degrade seriously when there are rich obstacles in the environment and/or when the communication users (CUs) and sensing targets are located far apart.

Recently, motivated by BS cooperation in communication \cite{Gesbert2010Multi} and distributed multiple-input multiple-out (MIMO) radar sensing \cite{haimovich2007mimo}, network ISAC (or perceptive mobile networks in \cite{zhang2021perceptive}) has attracted growing research interests \cite{huang2022coordinated,behdad2022power,Shi2022Device, zhang2021perceptive,wang2020constrained}
to resolve the above issues. On one hand, cooperative BSs can coordinate their transmission to better manage the inter-cell interference for enhancing the communication data rates at CUs \cite{Gesbert2010Multi}. On the other hand, different BSs can cooperatively sense targeted objects and environment from different angles, in which the waveform diversity gains can be exploited to enhance the sensing accuracy and resolution \cite{haimovich2007mimo}. However, due to the involvement of both coordinated multi-cell communication and distributed MIMO radar, how to properly design the transmit strategies at these BSs for balancing the sensing and communication performance tradeoffs is a challenging task.

In the literature, there have been a handful of prior works \cite{huang2022coordinated,behdad2022power,Shi2022Device,zhang2021perceptive,wang2020constrained} studying network ISAC. For instance, the authors in \cite{huang2022coordinated} studied a single-antenna network ISAC system, in which different BSs jointly optimize their transmit power control to minimize the total transmit power consumption,  while ensuring the individual signal-to-interference-plus-noise ratio (SINR) constraints at their respectively served CUs and the estimation accuracy or Cram{\' e}r-Rao bound (CRB) constraint for localizing one target. The work \cite{behdad2022power} considered cell-free massive MIMO for network ISAC with regularized zero-forcing beamforming, in which the BSs jointly optimize their transmit power control to maximize the sensing signal-to-noise ratio (SNR) while ensuring the communication SINR constraints at CUs. Furthermore, \cite{Shi2022Device} considered the device-free multi-target localization in a cooperative orthogonal frequency division multiplexing (OFDM) network, and \cite{wang2020constrained} investigated the network ISAC assisted by unmanned aerial vehicles (UAVs). Despite such research progress, how to design the {\it transmit beamforming/precoding strategies} for {\it multi-antenna} network ISAC systems for optimally balancing the sensing and communication performance tradeoffs is still not well investigated, thus motivating our investigation in this work.

This paper studies a multi-antenna network ISAC system, in which a set of multi-antenna BSs employ the coordinated transmit beamforming to serve their respectively associated CUs, and at the same time reuse the reflected information signals to perform target detection. Our main results are listed as follows.
\begin{itemize}
\item We consider two target detection scenarios depending on the time synchronization among BSs. In Scenario I, these BSs are all synchronized in time, and thus can exploit the target-reflected signals over both the direct links (from each BS to target to itself) and the cross links (from each BS to target to other BSs) for joint detection. In Scenario II, these BSs are not synchronized, and thus can only utilize the target-reflected signals over their direct links for joint detection.

\item  For each of the two scenarios, we apply the  likelihood ratio test for detection, and accordingly derive the detection probability under a specific false alarm probability at any given target location. The detection probability is shown to be monotonically increasing with respect to the total received reflection-signal power (over the utilized links for each scenario) at the BSs.

\item Based on the derivation in each scenario, we propose to optimize the coordinated transmit beamforming at the BSs to maximize the minimum detection probability (or equivalently the total received reflection-signal power) over a particular target area, while ensuring the minimum SINR constraints at the CUs, subject to the maximum transmit power constraints at the BSs. We use the SDR technique to obtain high-quality solutions to the formulated non-convex problems.

\item Finally, we provide numerical results to validate the performance of our proposed designs as compared to the benchmark schemes based on the communication design. It is shown that for each scenario, the proposed design achieves higher detection probability than the benchmark. It is also shown that the time synchronization among BSs in Scenario I is beneficial in enhancing the detection performance by exploiting more reflected signal paths over both direct and cross target-reflection links.
\end{itemize}

{\textit{Notation:}} Vectors and matrices are denoted by boldface lowercase and uppercase letters, respectively. $\bf{I}$ denotes an identity matrix with appropriate dimension. $\mathbb E (\cdot)$ denotes the statistical expectation. For a scalar $a$, $\left| a \right|$ denotes its absolute value. For a vector $\bf v$, $\left\| {\bf{v}} \right\|$ denotes its Euclidean norm. For a matrix ${\bf M}$ of arbitrary dimension, ${\bf M}^T$ and ${\bf M}^H$ denote its transpose and conjugate transpose, respectively. ${\mathbb C}^{x \times y}$ denotes the space of $x \times y$ complex matrices. $\cal{N}({\bf x}, {\bf Y})$ and $\cal{CN}({\bf x}, {\bf Y})$ denote the real Gaussian and the circularly symmetric complex Gaussian (CSCG) distributions with mean vector ${\bf x}$ and covariance matrix ${\bf{Y}}$, respectively, and ``$ \sim $" means ``distributed as". $Q(\cdot)$ denotes the Q-function.

\begin{figure}[ht]
\centering
    \includegraphics[width=7cm]{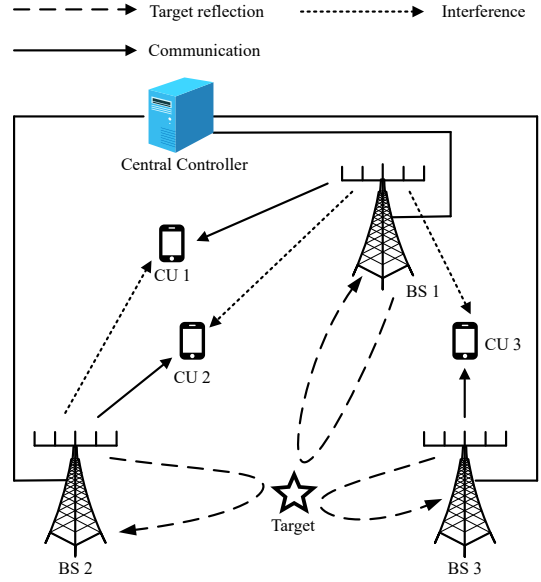}
\caption{An example multi-antenna network ISAC system model with three BSs/CUs.}
\label{fig:1}
\end{figure}

\section{System Model}


We consider a multi-antenna network ISAC system as shown in Fig. \ref{fig:1}, which consists of $K$ BSs each with $N_t>1$ transmit and $N_r>1$ receive antennas, as well as $K$ single-antenna CUs. Let $ {\cal K} \triangleq \left\{ {1, \ldots ,K} \right\}$ denote the set of BSs or CUs. In this system, the BSs send individual messages to their respectively associated CUs. At the same time, the BSs receive and properly process the reflected information signals and then send them to a central controller (CC) for joint target detection. As such, the multi-antenna network ISAC system unifies the multi-antenna interference channel for communication and the distributed MIMO radar detection, as will be detailed next. Specifically, we focus on the ISAC transmission over a particular block with duration $T$ that consists of $N$ symbols, where $T = N T_c$ with $T_c$ denoting the  duration of each symbol. Here, $T$ or $N$ is assumed to be sufficiently large for the ease of analysis. Let $\mathcal T \triangleq (0,T]$ denote the ISAC period of interest, and $\mathcal N \triangleq \{1, \ldots, N\}$  the set of symbols.

First, we consider the communication from the BSs to the CUs, in which the coordinated transmit beamforming is employed at these BSs. Let $\bar s_k\left(t\right)$ denote the information signal sent by BS $k$ for CU $k$ at time $t\in \mathcal T$, and $s_k[n]$ denote the sampled signal at each symbol $n\in \mathcal N$. Here, ${{{s}}_i}[n]$'s are assumed to be independent and identically distributed (i.i.d.) random variables with zero mean and unit variance. Let ${{\bf{w}}_k} \in {{\mathbb C}^{N_t\times 1}}$ denote the transmit beamforming vector by each BS $k \in \mathcal K$, and ${{\bf{h}}_{k,i}}\in \mathbb{C}^ {N_t\times 1}$ denote the channel vector from BS $i\in \mathcal K$ to CU $k\in \mathcal K$. Then the received signal at CU $k \in \mathcal K$ in symbol $n \in \mathcal N$ is
\begin{align}
{y_k}[ n ] &= {\bf{h}}_{k,k}^H{{\bf{w}}_k}{s_k}[ n ] + \sum\limits_{{i \in \cal K},i\neq k} {{\bf{h}}_{k,i}^H{{\bf{w}}_i}{s_i}[ n ]} + {z_k}[ n ],
\end{align}
where ${z_k}[ n ]  \sim  {\cal{CN}} \left( {0,\sigma ^2_c} \right)$ denotes the noise at the receiver of CU $k$, with $\sigma_c^2$ denoting the noise power. The received SINR at CU $k \in \mathcal K $ is given by
\begin{align}
\gamma _k = \frac{{{{\left| {{\bf{h}}_{k,k}^H{{\bf{w}}_k}} \right|}^2}}}{{\sum\limits_{{i \in \cal K},i\neq k} {{{\left| {{\bf{h}}_{k,i}^H{{\bf{w}}_i}} \right|}^2}}  + {\sigma ^2_c}}}.
\end{align}

Next, we consider the distributed MIMO radar detection by the $K$ BSs via reusing the information signals $\{\bar s_k(t)\}$. Let $(x_k, y_k)$ denote the location of each BS $k\in\cal{K}$. Suppose that there is one target present at location $(x_0,y_0)$, for which the target angle with respect to BS $k$ is denoted by $\theta_k$. Let ${{\bf{a}}_{t,k}}\left( {{\theta _k}} \right) \in \mathbb{C}^ {N_t\times 1} $ and ${\bf{a}}_{r,k}\left( {{\theta _k}} \right) \in \mathbb{C}^ {N_r\times 1} $ denote the transmit and receive steering vectors at BS $k\in\mathcal K$, respectively, where ${{\left\| {{{\bf{a}}_{t,k}}({\theta _k})} \right\|} \mathord{\left/{\vphantom {{\left\| {{{\bf{a}}_{t,k}}({\theta _k})} \right\|} {{N_t}}}} \right.\kern-\nulldelimiterspace} {{N_t}}} = {{\left\| {{{\bf{a}}_{t,k}}({\theta _k})} \right\|} \mathord{\left/{\vphantom {{\left\| {{{\bf{a}}_{t,k}}({\theta _k})} \right\|} {{N_r}}}} \right.\kern-\nulldelimiterspace} {{N_r}}} = 1$ is assumed without loss of generality. In the special case with uniform linear arrays (ULAs) deployed at BSs, we have ${{\bf{a}}_{t,k}}\left( {{\theta _k}} \right) = {\left[ {1,{e^{j2\pi \frac{{{d_a}}}{\lambda }\sin \left( {{\theta _k}} \right)}}, \ldots ,{e^{j2\pi \frac{{{d_a}}}{\lambda }\left( {N_t - 1} \right)\sin \left( {{\theta _k}} \right)}}} \right]^T}$ and ${{\bf{a}}_{r,k}}({\theta _k}){\rm{  = }}{\left[ {1,{e^{j2\pi \frac{{{d_a}}}{\lambda }\sin \left( {{\theta _k}} \right)}}, \ldots ,{e^{j2\pi \frac{{{d_a}}}{\lambda }\left( {{N_r} - 1} \right)\sin \left( {{\theta _k}} \right)}}} \right]^T}$ , where $j =\sqrt{-1}$, and $d_a$ and $\lambda$ denote the antenna spacing and wavelength, respectively. Let ${{\bf{H}}_{k,i}} = {{\hat \zeta }_{k,i}}{\bf{a}}_{r,k}\left( {{\theta _k}} \right){{\bf{a}}_{t,i}^T}\left( {{\theta _i}} \right) \in \mathbb{C}^ {N_r\times N_t} $ denote the target response matrix from BS $i$ to target to BS $k$, in which ${{\hat \zeta }_{k,i}} = \sqrt {{\beta _{k,i}}} {\zeta _{k,i}}$ is the reflection coefficient incorporating both the radar cross section (RCS) ${\zeta _{k,i}}$ and the round-trip path loss ${\beta _{k,i}}$. Specifically, we assume that ${\beta _{k,i}} = \kappa^2 { {\frac{{d_{\rm ref}^4}}{{{d_{k}^2}{d_{i}^2}}}} }$, where $\kappa$ denotes path loss at the reference distance $d_{\rm ref}$, and ${d_{k}} = \sqrt {({x_k} - {x_0}){}^2 + ({y_k} - {y_0}){}^2} $ denotes the distance between the target and BS $k$. We also assume an isotropic target with uniform RCS, i.e., ${\zeta _{k,i}}=\zeta , \forall i,k \in \cal K$. Under the above consideration, the received target-reflection signal at BS $k$ is given by
\begin{align}
{{\bf{r}}_k}\left( t \right) = \sum\limits_{i \in \mathcal{K}} {{{\bf{H}}_{k,i}}{{\bf{w}}_i}{\bar s_i}\left( {t - {\tau _{k,i}}} \right)}  + {{ \bf{\bar z}}_k}\left( t \right), \label{recsig}
\end{align}
where ${\bf{\bar z}}_k(t) \sim  {\cal{CN}} \left( {0,\sigma ^2_d{\bf{I}}} \right)$ denotes the noise at the receiver of BS $k$, and ${\tau_{k,i}} = \frac{1}{c}({d_{k}} + {d_{i}})$ denotes the transmission delay form BS $i$ to target to BS $k$, with $c$ denoting the speed of light. Without loss of generality, we assume that the information waveform has a normalized power over block $\mathcal T$, i.e., $\frac{1}{T}\int_\mathcal T {{{\left| {{\bar s_i}\left( t \right)} \right|}^2}dt}  = 1$. Furthermore, notice that $s_k(t)$'s are with zero mean and independent over different CUs and different time. As $T$ is sufficiently large, we have $ \frac{1}{T}\int_\mathcal T {{\bar s_i}\left( t \right)\bar s_k^*\left( {t - \tau } \right)dt}  = 0, \forall \tau, i\neq k$, and $\frac{1}{T} \int_\mathcal T \bar s_k(t) \bar s_k^* (t-\tau) dt = 0$ for any $k$ and $|\tau| \ge T_c$. Based on the received signals $\{{\bf r}_k(t)\}$ in (\ref{recsig}), the $K$ BSs jointly detect the existence of target. In particular, we consider two scenarios depending on  time synchronization among BSs.

{\bf Scenario \uppercase\expandafter{\romannumeral1}}: All the BSs are synchronized in time, such that the target-reflected signals over both direct links (i.e., ${{{\bf{H}}_{k,k}}{{\bf{w}}_k}{\bar s_k}\left( {t - {\tau _{k,k}}} \right)}$ from each BS $k$ to target to itself) and cross links (i.e., ${{{\bf{H}}_{k,i}}{{\bf{w}}_i}{\bar s_i}\left( {t - {\tau _{k,i}}} \right)}$ from other BSs $i$'s to target to BS $k$, $\forall i\neq k$) can be exploited for joint detection. Towards this end, each BS $k$ can perform the matched filtering (MF) processing based on ${{\bf{r}}_k}(t)$ by using $\bar s_i(t)$'s and delay $\tau_{k,i}$. Accordingly, the processed signal based on $\bar{s}_i(t)$ is 
\begin{align}
{{\bf d}_{k,i}} &=\frac{1}{T}\int_\mathcal T {{{\bf{r}}_k}\left( t \right)\bar s_i^*\left( {t - {\tau _{k,i}}} \right)dt} \nonumber \\ &= \underbrace {\frac{1}{T}\int_\mathcal T {{{\bf{H}}_{k,i}}{{\bf{w}}_i}{{\left| {\bar s_i^*\left( {t - {\tau _{k,i}}} \right)} \right|}^2}dt}  }_{\rm desired~signal} \nonumber \\
&~~~~+ \underbrace { \frac{1}{T}\int_\mathcal T {{{\bf{z}}_k}\left( t \right)\bar s_i^*\left( {t - {\tau _{k,i}}} \right)dt}  }_{\rm filtered ~noise}  = {{\bf{H}}_{k,i}}{{\bf{w}}_i} + {{\hat {\bf z}}_{k,i}},
\label{mod:1}
\end{align}
where ${\hat {\bf z}_{k,i}} \sim {\cal CN}\left({0,{\sigma ^2_d}{\bf I}} \right)$ denotes the equivalent noise after MF processing. After  obtaining $\{{{\bf d}_{k,i}}\}_{i\in\mathcal{K}}$, each BS $k$ shares them to the CC, which then performs the joint radar detection based on ${{\bf d}_{k,i}}$'s, $\forall i, k\in\mathcal{K}$. In this scenario, the usable signal is accumulated as ${\bf{d}}_{\rm I} = {\left[ {{\bf{d}}_{1,1}^T, \ldots ,{\bf{d}}_{1,K}^T,\ldots, {\bf{d}}_{K,1}^T,\ldots ,{\bf{d}}_{K,K}^T} \right]^T} \in {{\mathbb C}^{{N_rK^2}}}$.



{\bf Scenario \uppercase\expandafter{\romannumeral2}}: In this scenario, the BSs are not synchronized, and thus each BS $k$ cannot know the transmission delay $\tau _{k,i}$ with another BS $i\neq k$. In this scenario, the BSs can only utilize the target-reflected signals over their direct links (${{{\bf{H}}_{k,k}}{{\bf{w}}_k}{\bar s_k}\left( {t - {\tau _{k,k}}} \right)}$, $\forall k\in\mathcal K$) for joint detection. After the MF processing similarly as in Scenario I, we have the processed signal as ${\bf{d}}_{\rm II} = {\left[ {{\bf{d}}_{1,1}^T,{\bf{d}}_{2,2}^T, \ldots ,{\bf{d}}_{K,K}^T} \right]^T} \in {{\mathbb C}^{{N_rK}}}$.




\section{Detection Probability at Given Target Location }

In this section, we derive the detection probability and the false alarm probability at a given target location $(x_0,y_0)$ for the two scenarios. 

\subsection{Scenario \uppercase\expandafter{\romannumeral1} with BSs Synchronization}\label{Sec:III-A}

To start with, we define two hypotheses for target detection, i.e., ${\cal H}_1$ when the target exists and ${\cal H}_0$ when the target does not exist. For notational convenience, define ${{\boldsymbol \alpha} _{k,i}} = {{\bf{H}}_{k,i}}{{\bf{w}}_i}$ as the reflected signal vector from BS $i$ to target to BS $k$ when the target exists, and the correspondingly accumulated signal vector as ${\boldsymbol{\alpha }}_{\rm I} = {\left[ {{{\boldsymbol{\alpha }}_{1,1}^{T}}, \ldots ,{{\boldsymbol{\alpha }}_{1,K}^{T}}, \ldots ,{{\boldsymbol{\alpha }}_{K,1}^{T}}, \ldots ,{{\boldsymbol{\alpha }}_{K,K}^{T}}} \right]^T} \in {{\mathbb C}^{{N_rK^2}}}$. Furthermore, define the noise vector in Scenario I as ${\bf{\hat z}}_{\rm I} = {\left[ {{{\bf {\hat z}}_{1,1}^{T}}, \ldots ,{{\bf{\hat z}}_{1,K}^{T}}, \ldots ,{{\bf{\hat z}}_{K,1}^{T}}, \ldots ,{{\bf{\hat z}}_{K,K}^{T}}} \right]^T} \in {{\mathbb C}^{{N_rK^2}}}$. Then, based on \eqref{mod:1}, we have the processed signals after the MF processing as
\begin{align}\label{eqn:H_1:H_0}
\left\{ {\begin{array}{*{20}{c}}
{{\cal H}_1}: {{\bf{d}}}_{\rm I} = {{\boldsymbol{\alpha}}}_{\rm I} + {{\hat {\bf z}}}_{\rm I},\\
{{\cal H}_0}: {{\bf{d}}}_{\rm I} =  {{\hat {\bf z}}}_{\rm I}.
\end{array}} \right.
\end{align}



Next, we use the likelihood ratio test for target detection. Based on \eqref{eqn:H_1:H_0}, the likelihood functions of vector ${{\bf{d}}}_{\rm I}$ under hypothesis ${{\cal H}_1}$ and ${{\cal H}_0}$ are respectively given by
\begin{align}
{p\left( {{\bf{d}}_{\rm I}|{{\cal H}_1}} \right)} &\!= \! {\frac{1}{{{\pi ^{{N_rK^2}}}{\sigma ^{2{N_rK^2}}_d}}}\exp \left( { \!- \frac{1}{{{\sigma ^2_d}}}{{\left( {{\bf{d}}_{\rm I} \!-\! {\boldsymbol{\alpha }}_{\rm I}} \right)}^H}\!\left( {{\bf{d}}_{\rm I} \!-\! {\boldsymbol{\alpha }_{\rm I}}} \right)} \right)}, \\
{p\left( {{\bf{d}}|{{\cal H}_0}} \right)} & = {\frac{1}{{{\pi ^{{N_rK^2}}}{\sigma ^{2{N_rK^2}}_d}}}\exp \left( { - \frac{1}{{{\sigma ^2_d}}}{{\bf{d }}_{\rm I}^H}{\bf{d }}_{\rm I}} \right)}.
\end{align}
Accordingly, the Neyman-Pearson (NP) detector is given by the likelihood ratio test:
\begin{align}
\ln  \frac{{p\left( {{\bf{d}}_{\rm I}|{{\cal H}_1}} \right)}}{{p\left( {{\bf{d}}_{\rm I}|{{\cal H}_0}} \right)}} 
= \frac{1}{{{\sigma_d ^2}}}\left( {2{\rm{Re}}\left( {{{\boldsymbol{\alpha }}_{\rm I}^H}{\bf{d}}}_{\rm I} \right) - {{\boldsymbol{\alpha }}_{\rm I}^H}{\boldsymbol{\alpha }}_{\rm I}} \right) \mathop \gtrless\limits_{{\cal H}_0}^{{\cal H}_1}  \delta,\label{mod:2:v2}
\end{align}
where $\delta$ denotes the threshold determined by the tolerated level of false alarm. Notice that ${{\boldsymbol{\alpha }}_{\rm I}^H}{\boldsymbol{\alpha }}_{\rm I}$ is given, and as a result, the detector in \eqref{mod:2:v2} can be equivalently simplified as
\begin{align}\label{detector}
T\left( {\bf{d}}_{\rm I} \right) = {\rm{Re}}\left( {{{\boldsymbol{\alpha }}_{\rm I}^H}{\bf{d}}_{\rm I}} \right) \mathop \gtrless\limits_{{\cal H}_0}^{{\cal H}_1} {\delta}^{'},
\end{align}
where $\delta^{'}$ denotes the threshold related to $T\left( {\bf{d}} _{\rm I}\right)$.







Then, we derive the distribution of $T\left( {\bf{d}}_{\rm I} \right)$. Towards this end, we consider $x={{{\boldsymbol{\alpha }}_{\rm I}^H}{\bf{d}}_{\rm I}}$, 
whose expectation and variance under hypothesis ${{\cal H}_1}$ and ${{\cal H}_0}$ are obtained as
\begin{align}
{\mathbb{E} }\left( {x|{{\cal H}_0}} \right) 
& = 0, \label{exph0} \\
{\mathbb{E} }\left( {x|{{\cal H}_1}} \right) & = {\cal E} _{\rm I} \triangleq \sum\limits_{k\in\mathcal K} {\sum\limits_{i\in\mathcal K} {{{\left\| {{{\bf{H}}_{k,i}}{{\bf{w}}_i}} \right\|}^2}} } \nonumber\\&~~~~~~=N_r\zeta^2\sum\limits_{k\in\mathcal K} {\sum\limits_{i \in\mathcal K} {{\beta _{k,i}}{{\left| {{\bf{a}}_{t,i}^T\left( {{\theta _i}} \right){{\bf{w}}_i}} \right|}^2}} }, \label{exph1} \\
{\mathop{\rm var}} \left( {x|{{\cal H}_0}} \right) &= {\mathop{\rm var}} \left( {x|{{\cal H}_1}} \right) 
= {\sigma_d ^2} {\cal E}_{\rm I}  . \label{varh0}
\end{align}
By combining (\ref{exph0}), (\ref{exph1}), and (\ref{varh0}), we have
\begin{align}
 \left\{ {\begin{array}{*{20}{c}}
x \sim{{\cal CN}\left( {0,{\sigma_d ^2}{\cal E}_{\rm I} } \right),{{\cal H}_0}},\\
x \sim{{\cal CN}\left( {{\cal E}_{\rm I},{\sigma_d ^2}{\cal E}_{\rm I} } \right),{{\cal H}_1}}.
\end{array}} \right.
\end{align}
As a result, for $T\left( {\bf{d}}_{\rm I} \right) = {\mathop{\rm Re}\nolimits} \left( x\right)$, it follows that
\begin{align}\label{distribution}
\left\{ {\begin{array}{*{20}{c}}
T\left( {\bf{d}}_{\rm I} \right) \sim {{\cal N}\left( {0,{{{{\sigma ^2_d}{\cal E}_{\rm I} } \mathord{\left/
 {\vphantom {{{\sigma ^2_d}} 2}} \right.
 \kern-\nulldelimiterspace} 2}}} \right),{{\cal H}_0}},\\
T\left( {\bf{d}}_{\rm I} \right) \sim {{\cal N}\left( {{\cal E}_{\rm I} ,{{{{\sigma ^2_d}{\cal E}_{\rm I} } \mathord{\left/
 {\vphantom {{{\sigma ^2_d}} 2}} \right.
 \kern-\nulldelimiterspace} 2}}} \right),{{\cal H}_1}}.
\end{array}} \right.
\end{align}

Finally, we obtain the detection probability under given false alarm probability. Based on \eqref{detector} and \eqref{distribution}, we have the detection probability $p_D^{\rm I}$ and the false alarm probability ${p^{\rm I}_{FA}}$ with respect to the detector threshold $\delta^{'}$ as
\begin{align}
{p^{\rm I}_D} &= Q\left( {(\delta^{'}-{\cal E}_{\rm I} ) \sqrt {\frac{2}{{\sigma _d^2{\cal E}_{\rm I} }}} } \right),\label{pD1} \\
{p^{\rm I}_{FA}} &= Q\left( {\delta^{'} \sqrt {\frac{2}{{\sigma _d^2{\cal E}_{\rm I} }}} } \right).\label{pFA1}
\end{align}
Based on \eqref{pFA1}, we have ${\delta^{'} \sqrt {\frac{2}{{\sigma _d^2{\cal E}_{\rm I} }}} } ={Q^{ - 1}}\left( {{p^{\rm I}_{FA}}} \right)$. By substituting this into \eqref{pD1}, we obtain the detection probability under given false alarm probability $p_{FA}^{\rm I}$ as
\begin{align}
{p_D^{\rm I}} =  Q\left( {{Q^{ - 1}}\left( {{p_{FA}^{\rm I}}} \right) - \sqrt {\frac{{2{\cal E}_{\rm I} }}{{\sigma _d^2}}} } \right).
\end{align}
It is observed that the detection probability ${p_D^{\rm I}}$ is monotonically increasing with respect to ${\cal E}_{\rm I} = N_r\zeta^2\sum\limits_{k\in\mathcal K} {\sum\limits_{i \in\mathcal K} {{\beta _{k,i}}{{\left| {{\bf{a}}_{t,i}^T\left( {{\theta _i}} \right){{\bf{w}}_i}} \right|}^2}} }$, which corresponds to the  total received reflection-signal power over both direct and cross reflection links.


\subsection{Scenario \uppercase\expandafter{\romannumeral2} without BSs Synchronization}\label{Sec:III-B}
Next, we consider Scenario II without synchronization among BSs. The detection probability in this scenario can be similarly derived as that in Scenario I, by replacing ${\bf{d}}_{\rm I}$ by ${\bf{d}}_{\rm II}$ and accordingly replacing ${\cal E}_{\rm I}$ in \eqref{exph1} by
\begin{align}
{\cal E}_{\rm II} =  \sum\limits_{k \in \mathcal K} { {{{\left\| {{{\bf{H}}_{k,k}}{{\bf{w}}_k}} \right\|}^2}} } = N_r\zeta^2\sum\limits_{k \in \mathcal K}  {{\beta _{k,k}}{{\left| {{\bf{a}}_{t,k}^T\left( {{\theta _k}} \right){{\bf{w}}_k}} \right|}^2}} .
\end{align}
Based on the similar derivation procedure as in Section \ref{Sec:III-A}, we have the detection probability ${p_D^{\rm II}}$ under a given false alarm probability ${p_{FA}^{\rm II}}$ as
\begin{align}
{p_D^{\rm II}} = Q\left( {{Q^{ - 1}}\left( {p_{FA}^{\rm II}} \right) - \sqrt {\frac{{2{\cal {E}_{\rm II}}}}{{\sigma _d^2}}} } \right). \label{eq:19}
\end{align}

It is observed from \eqref{eq:19} that the detection probability ${p_D^{\rm II}}$ is monotonically increasing with respect to ${\cal E}_{\rm II} = N_r\zeta^2\sum\limits_{k = 1}^K  {{\beta _{k,k}}{{\left| {{\bf{a}}_t^T\left( {{\theta _k}} \right){{\bf{w}}_k}} \right|}^2}}$, which corresponds to the total received reflection-signal power over the direct links only.

\section{SINR-constrained Detection Probability Maximization via Coordinated Transmit Beamforming}
In this section, we are interested in designing the coordinated transmit beamforming $\{{{{\bf{w}}_k}}\}$ to maximize the minimum detection probability over a particular target area with a given false alarm probability $p_{FA}$, subject to the minimum SINR requirement $\Gamma_k$ at each CU $k\in \cal K$, and the maximum power constraint $P_{\max}$ at each BS. In particular, let $\cal M$ denote the target area for detection. To facilitate the design, we take $M$ sample locations from $\cal M$, denoted by $(x_0^{(m)}, y_0^{(m)}), \forall m\in {\cal M} \triangleq \{1,\ldots, M\}$. For a potential target located at $(x_0^{(m)}, y_0^{(m)})$, we denote the target angle with respect to BS $i\in\mathcal K$ as $\theta_i^{(m)}$ and the round-trip path loss from BS $i\in\mathcal K$ to target to BS $k\in\mathcal K$ as ${\beta _{k,i}^{(m)}}$.

\subsection{Scenario \uppercase\expandafter{\romannumeral1} with BSs Synchronization}

First, we consider Scenario I with BSs Synchronization. In this scenario, for given target location $m$, it follows from \eqref{pFA1} that the detection probability is monotonically increasing with respect to the total reflection-signal power $\sum\limits_{k \in \mathcal K} {\sum\limits_{i \in \mathcal K} {{\beta^{(m)} _{k,i}}{{\left| {{\bf{a}}_{t,i}^T\left( {{\theta^{(m)}_i}} \right){{\bf{w}}_i}} \right|}^2}} }$. Therefore, maximizing the detection probability in this scenario is equivalent to maximizing $\sum\limits_{k \in \mathcal K} {\sum\limits_{i \in \mathcal K} {{\beta^{(m)} _{k,i}}{{\left| {{\bf{a}}_{t,i}^T\left( {{\theta^{(m)}_i}} \right){{\bf{w}}_i}} \right|}^2}} }$. Based on this observation, the SINR-constrained minimum detection probability  maximization over the given target area can be formulated as the following optimization problem:
\begin{subequations}
\begin{align}
\left( {\rm{P}}{\rm{1}} \right): \mathop {\max }\limits_{\left\{ {{{\bf{w}}_i}} \right\}} ~ &~ \mathop {\min }\limits_{m\in \cal{M}}  ~\sum\limits_{k\in \mathcal K} {\sum\limits_{i \in \mathcal K} {{\beta^{(m)} _{k,i}}{{\left| {{\bf{a}}_{t,i}^T\left( {{\theta^{(m)}_i}} \right){{\bf{w}}_i}} \right|}^2}} }   \label{prob:1} \\
~{\rm{s}}{\rm{.t}}{\rm{.}}&~~\frac{{{{\left| {{\bf{h}}_{k,k}^H{{\bf{w}}_k}} \right|}^2}}}{{\sum\limits_{{i \in \cal K},i\neq k} {{{\left| {{\bf{h}}_{k,i}^H{{\bf{w}}_i}} \right|}^2}}  + {\sigma ^2_c}}} \ge {\Gamma _k}, \forall k \in {\cal K} \label{prob:con:1} \\
&~~{\left\| {{{\bf{w}}_i}} \right\|^2} \le {P_{\max }},\forall i \in {\cal K}, \label{prob:con:2}
\end{align}
\end{subequations}
where (\ref{prob:con:1}) denotes the minimum SINR constraints at the CUs and (\ref{prob:con:2}) denotes the maximum transmit power constraints at the BSs. Notice that problem (P1) is non-convex due to the non-convex constraints in (\ref{prob:con:1}). In the following, we apply the SDR technique to solve this problem \cite{luo2010semidefinite}.


Towards this end, we introduce $t$ as an auxiliary variable, and define ${{\bf{W}}_k} = {{\bf{w}}_i}{\bf{w}}_i^H \succeq \bf{0}$ with ${\rm {rank}} ({\bf{W}}_i)=1$, $\forall i \in \cal K$. Furthermore, define ${\bf{A}}_i({\theta^{(m)} _i}) = {\bf{a}}_{t,i}^ * ({\theta^{(m)} _i}){\bf{a}}_{t,i}^T({\theta^{(m)} _i}),\forall i\in\mathcal{K},m\in\mathcal{M}$. Problem (P1) can be equivalently reformulated as
\begin{subequations}
\begin{align}
&\left( {\rm{P}}{\rm{1.1}} \right):\mathop {\max }\limits_{\left\{ {{{\bf{W}}_i}\succeq \bf{0}},t \right\}} ~t    \label{prob:1.1} \\
&~~~~{\rm{s}}{\rm{.t}}{\rm{.}}~ \sum\limits_{k \in \mathcal K} {\sum\limits_{i \in \mathcal K} {{\beta^{(m)} _{k,i}}{\rm{tr}}\left( {{{\bf{W}}_i}{\bf{A}}_i\left( {{\theta^{(m)} _i}} \right)} \right)} } \ge t, \forall m \in \cal M\\
&~~~~~~~~\sum\limits_{{i \in \cal K}, i \ne k} {{\rm{tr}}\left( {{{\bf{h}}_{k,i}}{\bf{h}}_{k,i}^H{{\bf{W}}_i}} \right)}  \nonumber \\
&~~~~~~~~+ {\sigma ^2_c} \le \frac{{{\rm{tr}}\left( {{{\bf{h}}_{k,k}}{\bf{h}}_{k,k}^H{{\bf{W}}_k}} \right)}}{{\Gamma _k}}, \forall k \in {\cal K} \label{prob:con:1.1} \\
&~~~~~~~~{\rm tr}\left( {{{\bf{W}}_i}} \right)   \le {P_{\max }},\forall i \in {\cal K} \label{prob:con:2.1}\\
&~~~~~~~~{\rm {rank}} ({\bf{W}}_i)=1, \forall i \in \cal K.\label{prob:con:3.1}
\end{align}
\end{subequations}
However, problem (P1.1) is still non-convex due to the rank-one constraints in \eqref{prob:con:3.1}. To tackle this issue, we drop these rank-one constraints and obtain the SDR version of (P1.1) as (SDR1.1), which is convex and can be optimally solved by standard convex optimization solvers such as CVX \cite{grant2014cvx}. Let $\{{{\bf{W}}_i^\star}\}$ and $t^\star$ denote the obtained optimal solution to problem (SDR1.1). In practice, $\{{{\bf{W}}_i^\star}\}$ may not be of rank one and thus are not feasible or optimal for problem (P1.1). Therefore, we need to implement additional steps to find the solution to (P1.1) and thus (P1).

Finally, we apply the Gaussian randomization to construct rank-one solutions to (P1.1). Let the eigenvalue decomposition (EVD) of ${{\bf{W}}_i^\star}$ be denoted by ${{\bf{W}}_i^\star} = {\bf{U}}_i{\bf{\Sigma}}_i{\bf{U}}_i^H, \forall i \in {\cal K}$. In each randomization $n$ among the $N_G$ realizations, we randomly generate ${\bf{r}}_i^{(n)} = \mathcal{CN}({\bf{0}},{\bf{I}})$ and accordingly have ${{\bf{u}}_i^{(n)}} = {\bf{U}}_i{\bf{\Sigma}}_i^{1/2}{\bf{r}}_i^{(n)}, \forall i \in {\cal K}$. Then we set ${\bf{w}}_i^{(n)} = \sqrt{p_i^{(n)}} {{{\bf{u}}_i^{(n)}} \mathord{\left/
{\vphantom {{{\bf{u}}_i^{(n)}} {\left\| {{\bf{u}}_i^{(n)}} \right\|}}} \right.
\kern-\nulldelimiterspace} {\left\| {{\bf{u}}_i^{(n)}} \right\|}}$ and ${{\bf{W}}_i} = {\bf{w}}_i^{(n)}{( {{\bf{w}}_i^{(n)}} )^H}$, $\forall i \in {\cal K}$, where $\{p_i^{(n)}\}$ are optimization variables denoting the transmit powers at the $K$ BSs. Here, $\{p_i^{(n)}\}$ can be determined by substituting $\{{\bf{w}}_i^{(n)}\}$ into (P1) and solving the resultant linear program. By implementing the above procedures $N_G$ times, we choose the best $\{{\bf{w}}_i^{(n)}\}$ and $\{{\bf{W}}_i^{(n)}\}$ as the obtained solutions to (P1) and (P1.1), respectively.

\subsection{Scenario \uppercase\expandafter{\romannumeral2} without BSs Synchronization}

Next, we consider Scenario II without BSs synchronization. In this scenario, the SINR-constrained minimum detection probability maximization problem is formulated as problem (P2) in the following, which is similar as problem (P1) by replacing
$\sum\limits_{k \in \mathcal K} {\sum\limits_{i \in \mathcal K} {{\beta^{(m)} _{k,i}}{{\left| {{\bf{a}}_{t,i}^T\left( {{\theta^{(m)}_i}} \right){{\bf{w}}_i}} \right|}^2}} }$ in $\cal{E}_{\rm{I}}$ as $\sum\limits_{k \in \mathcal K} {{\beta^{(m)} _{k,k}}{{\left| {{\bf{a}}_{t,k}^T\left( {{\theta^{(m)}_k}} \right){{\bf{w}}_k}} \right|}^2}}$ in $\cal{E}_{\rm{II}}$.
\begin{align*}
\left( {\rm{P}}{\rm{2}} \right): \mathop {\max }\limits_{\left\{ {{{\bf{w}}_i}} \right\}} ~ &~ \mathop {\min }\limits_{m\in \cal{M}}  ~\sum\limits_{k \in \mathcal K} {{\beta^{(m)} _{k,k}}{{\left| {{\bf{a}}_{t,k}^T\left( {{\theta^{(m)}_k}} \right){{\bf{w}}_k}} \right|}^2}}  \\
~{\rm{s}}{\rm{.t}}{\rm{.}}&~~\eqref{prob:con:1}~{\text{and}}~\eqref{prob:con:2}.
\end{align*}
As problem (P2) has a similar structure as (P1), it can also be solved based on the SDR together with the Gaussian randomization. Therefore, the detailed solution procedure is omitted here for brevity.

\section{Numerical Results}
In this section, we provide numerical results to validate the detection performance of our proposed coordinated transmit beamforming designs in the multi-antenna network ISAC system, as compared to the following benchmark scheme based on communication design only.
\begin{itemize}
\item {\bf Communication design}: First, we design the transmit beamforming vectors $\{{{{\bf{w}}_i}}\}$ to minimize the total transmit power $\sum_{i=1}^K \|{\bf{w}}_i\|^2$ while ensuring the SINR constraints in \eqref{prob:con:1}, for which the obtained transmit beamformers are denoted by $\{{\bf{\bar{w}}}_i\}$. Then, we scale them as ${\bf{\hat{w}}}_i = \hat{p} {\bf{\bar{w}}}_i, \forall i \in \mathcal{K}$, where $\hat{p} > 0$ is set as the maximum value provided that the maximum transmit power constraints in \eqref{prob:con:2} are ensured.
\end{itemize}

In the simulation, each BS is deployed with a ULA with half wavelength spacing between antennas, and Rician fading is considered for communication. The noise powers are set as $\sigma^2_c = -84$ dBm and $\sigma^2_d = -102$ dBm after the MF processing with the ISAC transmission duration being $N=64$. The SINR constraints at CUs are set to be identical, i.e.,  $\Gamma_k=\Gamma$, $\forall k \in \cal K$. Furthermore, suppose that there are $K=3$ BSs and CUs in the system. The coordinates of the three BSs are $\left( {-60 \text{m},0} \right)$, $\left({60 \text{m},0}\right)$, and $\left( {0,60 \text{m}} \right)$, those of the three CUs are  $\left( { - 10 \text{m},0} \right)$, $\left( {  10 \text{m},0} \right)$, and $\left( {0,10 \text{m}} \right)$. The numbers of transmit and receive antennas at BSs are $N_t=N_r=N_a$. In addition, the target area is set as a square with area of $3 \times 3=9$ square meters and with the center at origin, and there are $M=9$ sample locations uniformly distributed in the target area.


\begin{figure}[!t]
\centering
    \includegraphics[width=9cm]{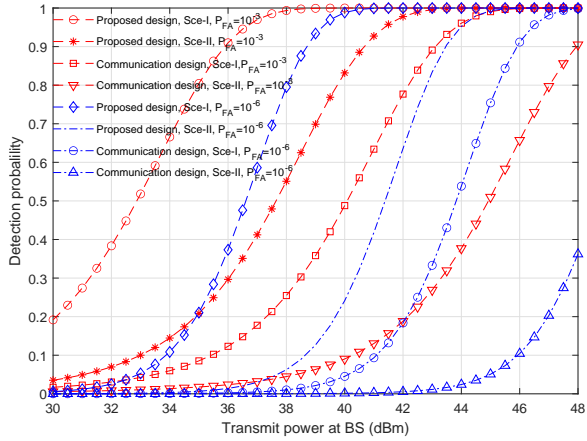}
\caption{The detection probability versus the maximum transmit power $P_{\max}$ at BS, where $N_a=32$, $\Gamma =10$ dB.}  
\label{fig:2}
\end{figure}

Fig. \ref{fig:2} shows the detection probability $p_D$ versus the maximum transmit power constraint $P_{\max}$ at the BSs, in which the SINR constraints at CUs are set as $\Gamma = 10$ dB and the number of antennas is $N_a=32$. In this figure, we consider two different false alarm probabilities, i.e, $p_{FA}=10^{-3}$ and $10^{-6}$, respectively. It is observed that for all the schemes, the detection probability increases towards one as the transmit power becomes large. The proposed design with Scenario I is observed to achieve the highest detection probability. This is due to the fact that in this scheme, the transmit beamforming vectors are properly designed to balance the communication and detection performances, and both direct and cross signal-reflection paths are exploited for joint detection. Furthermore, for each scenario, the proposed design is observed to outperform the communication design. This shows the benefit of our proposed design for multi-antenna network ISAC.

\begin{figure}[!t]
\centering
    \includegraphics[width=9cm]{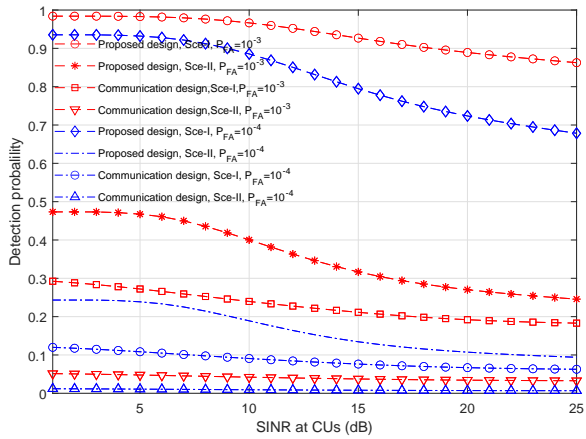}
\caption{The detection probability versus the SINR $\Gamma$ at CUs, where  $N_a=16$ and $P_{\max}=42$ dBm.}
\label{fig:3}
\end{figure}

Fig. \ref{fig:3} shows the detection probability $p_D$ versus the SINR requirements $\Gamma$ at the CUs, in which the maximum transmit power is $P_{\max} = 42$ dBm and the number of antennas is $N_a=16$. In this figure, we consider two cases with $p_{FA}=10^{-3}$ and $10^{-4}$, respectively. It is observed that for all the schemes, the detection probability decreases as the SINR requirement becomes large. This is due to the fact that when the communication requirements become stringent, the BSs need to steer the transmit beamformers towards the CUs, thus leading to less power towards the target location. Furthermore, it is observed that under given $p_{FA}$ for each scenario, the proposed design achieves much higher detection probabilities than the communication design, thanks to the coordinated transmit beamforming design that jointly considers the communication and target detection.


\section{Conclusion}
This paper studied the joint multi-cell communication and distributed MIMO radar detection in a multi-antenna network ISAC system, in which the BSs reuse their information signals for target detection. We considered two joint detection scenarios with and without time synchronization among the BSs, for which the detection probability and the false alarm probability are derived in closed form. Accordingly, we developed the coordinated transmit beamforming design to maximize the detection probability while ensuring the SINR constraints at CUs for communication. Numerical results show that the proposed coordinated transmit beamforming design together with the time synchronization among BSs achieves the best detection and communication performances.


\bibliographystyle{IEEEtran}
\bibliography{IEEEabrv,myref}

\end{document}